\documentclass[twocolumn]{aastex631}

\usepackage{graphicx} 
\usepackage{hyperref}
\usepackage{CJKutf8}
\usepackage{booktabs}
\usepackage{amsmath}

\begin{document}    

\title{You Only Stack Once (YOSO): A Motion-Filtered, Deep-Learning Framework for Detecting Faint Moving Sources
     
} 

\newcommand{\linccfw}{LSST Interdisciplinary Network for Collaboration and Computing, Tucson, USA}
\newcommand{\diracuw}{Dept. of Astronomy \& the DiRAC Institute, University of Washington, Seattle, USA}
\newcommand{\nau}{Department of Astronomy and Planetary Science, Northern Arizona University, Flagstaff, USA}

\author[0009-0003-4601-8556]{Nitya Pandey}
\email{npandey@das.uchile.cl}
\affiliation{Departamento de Astronomía, Universidad de Chile, Camino del Observatorio 1515, Las Condes, Santiago, Chile}
\affiliation{Centro de Excelencia en Astrofísica y Tecnologías Afines (CATA), Chile}

\author[0000-0002-0786-7307]{César Fuentes}
\email{cfuentes@das.uchile.cl}
\affiliation{Departamento de Astronomía, Universidad de Chile, Camino del Observatorio 1515, Las Condes, Santiago, Chile}
\affiliation{Centro de Excelencia en Astrofísica y Tecnologías Afines (CATA), Chile}

\author[0000-0003-0743-9422]{Pedro Bernardinelli}
\affiliation{\diracuw}

\author{Valeria Frías}
\affiliation{Facultad de Ciencias Físicas y Matemáticas (FCFM), University of Chile, Beauchef 850, 851, Santiago, Chile}

\author[0000-0001-7335-1715]{Colin Orion Chandler}
\affiliation{\diracuw}
\affiliation{\linccfw}

\author[0000-0003-4580-3790]{David E. Trilling}
\affiliation{\nau}

\author[0000-0002-1139-4880]{Matthew J. Holman}
\affiliation{Harvard-Smithsonian Center for Astrophysics, 60 Garden Street, MS 51, Cambridge, MA 02138, USA}

\author[0000-0002-7712-6678]{Steven Stetzler}
\affiliation{Jet Propulsion Laboratory, California Institute of Technology, 4800 Oak Grove Dr., Pasadena, CA 91109 USA}

\author[0000-0003-4051-2003]{Dallin Spencer}
\affiliation{Brigham Young University, Department of Physics and Astronomy, N283 ESC, Provo, UT 84602, USA}

\author[0000-0001-7737-6784]{Hsing~Wen~Lin (\begin{CJK*}{UTF8}{bkai}林省文\end{CJK*})}
\affiliation{Department of Physics, University of Michigan, Ann Arbor, MI 48109, USA}
\affiliation{Michigan Institute for Data and AI in Society, University of Michigan, Ann Arbor, MI 48109, USA}

\author[0000-0002-6514-318X]{Luis E. Salazar Manzano}
\affiliation{Department of Astronomy, University of Michigan, Ann Arbor, MI 48109, USA}

\author[0000-0003-1080-9770]{Darin Ragozzine} 
\affiliation{Brigham Young University, Department of Physics and Astronomy, N283 ESC, Provo, UT 84602, USA}

\author[0000-0001-6350-807X]{Ryder Strauss} 
\affiliation{\nau}

\author[0000-0003-1996-9252]{Mario Jurić}
\affiliation{\diracuw}

\author[0000-0001-5576-8189]{Andrew J. Connolly}
\affiliation{eScience Institute, Department of Astronomy, University of Washington, Seattle, WA 98195-1580, USA}

\author[0000-0002-7895-4344]{Hayden Smotherman}
\affiliation{\diracuw}

\author[0000-0003-3145-8682]{Scott S. Sheppard}
\affiliation{Earth and Planets Laboratory, Carnegie Institution for Science, Washington, DC 20015}

\author[0000-0003-4827-5049]{Kevin Napier}
\affiliation{Department of Physics, University of Michigan, Ann Arbor, MI 48109, USA}

\begin{abstract}
We present \texttt{You Only Stack Once (YOSO)}, an automated pipeline designed to detect faint, slow-moving Solar System objects in wide-field astronomical surveys. The pipeline integrates a novel Gaussian Motion Filter (GMoF) that operates at the pixel level to enhance signal-to-noise for objects exhibiting a range of apparent rates of motion. Unlike conventional shift-and-stack methods, which rely on discrete velocity trials, GMoF amplifies trails while suppressing random noise and static background features. Applied to a subset of DEEP observations from the Dark Energy Camera, YOSO discovered 45 out of 73 previously detected objects, as well as 11 new TNOs. It also discovered 216 objects in the near Solar System. Although alternative shift-and-stack methods are sensitive to objects about 0.88 magnitudes fainter, YOSO's false positive rate is extremely low, since it detects only sources that exhibit a trail and are consistent with a point source when shifted at the right rate. We show how this method can be deployed on large surveys like LSST, and be adapted for other domains that require motion-based signal enhancement, including exoplanet imaging through Angular Differential Imaging (ADI), near-Earth object (NEO) detection for missions like NEO Surveyor. YOSO thus provides a versatile, scalable approach for extracting faint, motion-dependent signals in the era of data-intensive astronomy.
\end{abstract}


\section{Introduction}
Despite their significance for understanding planetesimal formation processes and Solar System dynamical evolution, TNOs are challenging to detect due to their faint magnitudes (often $m > 22$). Since TNOs reflect sunlight, their visibility depends on their albedo and distance, making deep searches necessary for their detection. The traditional approach to detecting moving objects in astronomical images relies on identifying their parallactic motion. Over the course of a night, these objects move linearly across the sky. Distant objects produce shorter trails, whereas nearby and fast-moving ones, such as Near Earth Objects (NEOs), leave longer streaks. This characteristic motion provides a distinctive signature that can be exploited for their detection \citep{edgeworth1943evolution, kuiper1951origin}.

A well-established technique for detecting faint moving objects is the shift-and-stack method \citep{tyson1992limits}, and its more advanced variants, such as the Kernel-Based Moving Object Detection (KBMOD) \citep{whidden2019fast, smotherman2021sifting}, involve systematically applying trial-and-error adjustments to align images in a way that compensates for the parallactic motion of the objects being observed. The goal is to align the images to the point where a moving object appears as a stationary point source in the combined image. The deepest digital tracking to date has been done by \cite{bernstein2004size}, reached mean magnitude $\approx$ 28.3 (diameter $\approx$ 25km) using Advanced Camera for Surveys (ACS) aboard the Hubble Space Telescope. One of the main drawbacks is the potential for false positives. As the speed of the moving objects or the duration of observation increases, the number of required trials also grows. This increases the likelihood of mistakenly identifying stationary sources as moving objects, leading to false-positive detections.

In this study, we present a novel deep learning-based method for detecting TNOs. Our methodology leverages Convolution Neural Networks (CNNs) to identify TNO trails in combined images, offering an efficient alternative to other techniques. Over the past decade, CNNs have become powerful tools for image-based detection and classification, demonstrating remarkable performance across a wide range of astronomical applications, such as in exoplanet discovery \citep{ iglesias2023one, shallue2018identifying}, stellar astronomy \citep{ostdiek2020cataloging, li2025machine}, anomaly detection \citep{mesarcik2023road}. Within the context of Solar System science, CNN-based approaches have also been successfully applied to the detection of moving objects, such as streak identification in wide-field surveys \citep{duev2019deepstreaks} and more general machine-learning frameworks for moving-object detection \citep{fraser2025detecting}. 

The subsequent sections of the paper are organized as follows: first, in the section \ref{sec:data}, we provide an overview of the data employed in our study. Next, we detail our Method in the section \ref{sec:method}, outlining the algorithmic choices made during the analysis. Subsequently, in the results section \ref{results}, we present the efficiency of our method and the discovered objects. In the analysis section, we compare our detection parameters with those derived by the DEEP collaboration. Following that, in the discussion section \ref{discussion}, we critically evaluate the limitations of our method and propose potential avenues for improvement. Finally, in the conclusions section, we delve into the future prospects of AI-powered search algorithms, particularly their application in Solar System science and other surveys with limited brightness constraints.

\section{Data} \label{sec:data}
We considered data from the DECam Ecliptic Exploration Project (DEEP) (\citealt{trilling2024decam}, \citealt{trujillo2024decam}), the largest multiyear outer Solar System ecliptic survey to date. The survey had been allocated 46.5 nights to explore the trans-Neptunian Solar System using the Dark Energy Camera (DECam), a 3 square degree imager situated on the Blanco telescope at the Cerro Tololo Inter-American Observatory (CTIO, \citealt{2015AJ....150..150F}) in Chile. The survey aims to study small and faint TNOs down to 20~km in size, understand their shapes and properties based on their dynamic class, and to determine their orbits over three or more years more \citep{trilling2024decam, bernardinelli2024decam, trujillo2024decam}. 

The DEEP survey acquired \textit{VR}-band DECam exposures over four years, from 2019 to 2022. The \textit{VR} filter is suitable for the new object detection, because of its broad wavelength range. The complete DEEP survey was divided into four quadrants: \textit{A0}, \textit{A1}, \textit{B0}, and \textit{B1}, where  \textit{A} and \textit{B} represent the observing semesters, observed in the first half and the second half of the calendar year, respectively. Each of the \textit{A} and \textit{B} groups was subdivided into fields, as can be seen in the Table \ref{tab:decam_fields}. DECam Field of View has an approximately circular shape and a diameter of about 2.2 degrees, generating a total of 254 unique Field Nights (FNs) spanning a triangular pattern over the sky to account for the collective spreading of the orbit caused by the gravitational forces from nearby celestial bodies and influence the eccentricity variations \citep{trujillo2024decam}. Each FN contains a series of consecutive 120~s exposures, typically taken over 
a 4-hour period.

\begin{table}[ht]
    \centering
    \begin{tabular}{|c c c c c c|} 
     \hline
     Field & Night & RA [$^\circ$] & Dec [$^\circ$] & $N_\mathrm{exps}$ & Seeing \\ [0.2ex] 
     \hline
     B1b & 2020-10-15 & 352.863196 & -5.609639 & 103 & 1.46 \\ 
    
     B1h & 2021-09-08 & 356.235704 & -4.206055 & 70 & 1.44 \\ 
     
     B1i & 2021-10-03 & 355.136983 & -2.698944 & 100 & 1.17 \\
    
     B1j & 2021-09-30 & 354.436454 & -1.026361 & 85 & 1.52 \\
     \hline
    \end{tabular}
    \caption{The table summarizes the FNs we chose for the analysis. The first column shows the names of the field, and the next column shows which night that field was observed in UT. For this work, the FNs were chosen to incorporate a range of the length of stare ($N_\mathrm{exps}$) and the quality (seeing in arcseconds) of the data for that night.}
    \label{tab:decam_fields}
\end{table}

\begin{figure}[ht!]
\plotone{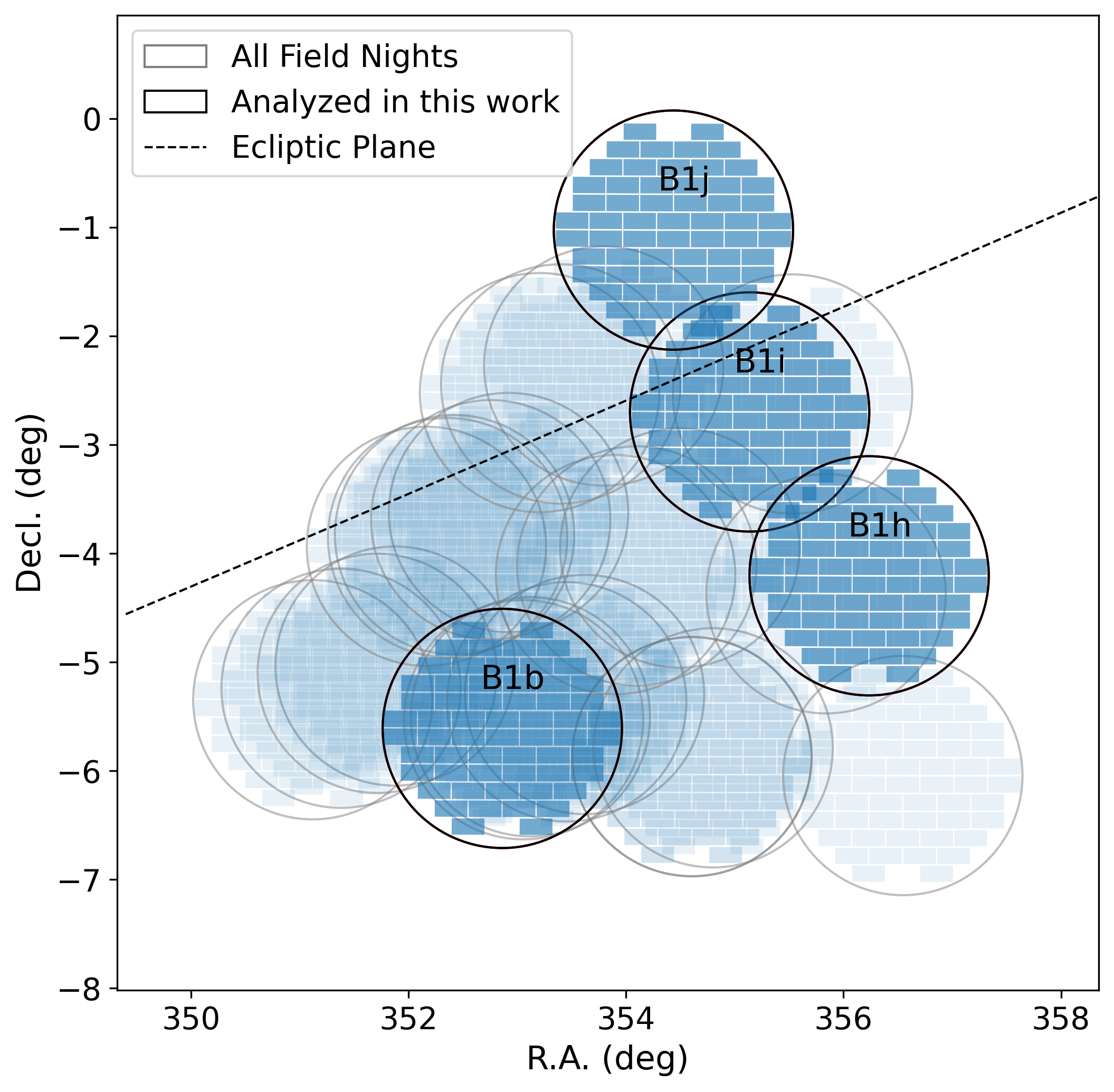}
\caption{On-sky positions of DECam field pointings for quadrant $B1$. The footprint consists of 29 field nights, covering approximately 30 square degrees. Each field received $\sim$3.8 hours of integration time with individual exposures of 120\,s. Field centers are defined by the R.A. and Dec.\ of the first exposure in each long stare. Of the 29 field nights, four have been analyzed by us to date, these are shown in dark blue patches with black circles. The dashed curve indicates the Solar System's ecliptic plane.
\label{fig:FN}}
\end{figure}

This work revisits an analysis of the $B1$ quadrant, previously processed using the shift-and-stack method by the DEEP collaborators \citep{smotherman2024decam,napier2024decam}. The $B1$ quadrant encompasses a total of 10 unique fields across 29 nights of observations, covering a footprint of approximately 30 square degrees (see Fig. \ref{fig:FN}). On average, each FN in $B1$ included 80 exposures. For instance, FN B1b, observed on October 15, 2020, consists of $\sim$~100 exposures of 120 seconds each, resulting in a total observation time of 3 hours and 34 minutes. 

From the 29 unique FNs in the $B1$ quadrant analyzed by \citet{smotherman2024decam}, we have selected four representative FNs that span a range of observational conditions. We restricted our analysis to the same DEEP survey quadrant in order to use the benchmark field-by-field magnitudes established in that work, allowing a direct comparison with our method’s limiting magnitude.

\subsection{Processing of DECam data} \label{sub:preprocessing}
For this study, we have used DECam’s \texttt{InstCal} images, which are photometrically and astrometrically calibrated by the DECam Community Pipeline, developed by \cite{valdes2014decam}, and \cite{decam-pipeline}. Our goal is to measure discovery efficiency as a function of magnitude. To accomplish this, we injected synthetic TNOs (or ``fakes'') with uniformly distributed magnitudes and orbits representative of the TNO population into the \texttt{InstCal} images. This approach ensures that we have a sufficient number of sources at various magnitudes and orbital configurations, allowing us to construct a robust efficiency curve.

For our chosen FNs  (table \ref{tab:decam_fields}), we generated over 20,000 unique synthetic objects around each FN center, with central coordinates in Right Ascension (RA) and Declination (Dec) and a radius of 1.1 degrees (Fig.~\ref{fig:FN}). For the entire synthetic population,
we assigned a magnitude $m_{VR}$ that uniformly ranges from  $19$ to $27$, where $19 \leq m_{VR} \leq 23.5$ are detectable in individual images, and the objects with magnitude $23.5 \leq m_{VR} \leq 27$ are beyond the detection threshold in single images.  

Generated synthetic objects were propagated throughout the epochs of the observation for each long stare. As an example, for FN~\texttt{\char96 B1b 20201015\char39}, which consists of 103 epochs, we generated a population of 5,000 synthetic objects at the first epoch based on the field’s central right ascension and declination. These objects were then propagated consistently across the remaining epochs. The process created an ephemeris of moving objects that span $30 < a < 80$ AU, $0^\circ < i < 180^\circ$, $0 < e < 0.99$, and $19 < m_{VR} < 27$. This moving object catalog is then implanted in each image, considering the seeing of each image, and CCD. 

After implantation, images went through six steps of our pre-processing pipeline. First, planted images went through the local background subtraction to eliminate the atmospheric variation from the images using the methods implemented in \texttt{photutils} \citep{bradley2020astropy}. Local background subtraction was performed using \texttt{Background2D}  with a mesh size of $100\times100$\, pixels and a $3\times3$ median filter to smooth large-scale gradients. Within each mesh, the background level was estimated through iterative $3\sigma$ clipping to exclude astrophysical sources, yielding a two-dimensional background model that was subtracted from the original image to produce flattened, background-corrected frames. This step was important for ensuring that the atmospheric variation across the images per FNs are almost the same, and to help in flagging the smallest variation of the faintest moving objects in the final combined images. After background subtraction, we registered all images to the one closest to the average pointing, as to minimize the shift for any given observation.
We then median combined these interpolated images into a "sky" to keep mostly stationary objects. 
This sky was then subtracted from each interpolated image using the ISIS package \citep{alard1998method}, yielding difference images produced by the subtraction pipeline. ISIS is based on the Alard–Lupton optimal image subtraction technique, which models spatially varying convolution kernels to transform a reference image so that its point-spread function (PSF) matches that of the target image before subtraction. By minimizing residuals arising from differences in seeing and background, this method produces high-quality difference images. After applying ISIS, the resulting subtracted frames were free of stationary sources, leaving mostly only moving and varying objects. At this stage, the images were ready to be combined.

\section{A new detection technique} 
\label{sec:method}
With only varying sources in the images, methods like shift-and-stack explore all plausible velocity vectors, average the shifted images, and search for point sources on them. This leads to a large number of false positives that grows with the number of trials.

We address this limitation by trying to remove the need to shift at the right rate for every moving object in the field, looking at ways of combining the unshifted subtracted images only once. 
For shift-and-stack methods the average statistic makes sense as it mimics taking a long stare at the field while tracking a certain motion across the sky. However, combining images at the wrong rate diffuses a moving object signal. We first experimented with higher-order statistics, such as kurtosis and skewness, which were effective at enhancing SNR for over the mean in the unshifted images but only enhanced the detection depth slightly (0.3 magnitudes better than single image detection after combining 100 images).

We introduce an alternative combination method 
using a new statistic that takes into consideration that a moving object leaves a characteristic signal on a pixel lightcurve. In this "stacked" image, pixels that have been traversed by a moving object are enhanced, appearing as  visible linear trails that can then be found using machine learning methods.

\begin{figure}[ht!]
    \centering  
    \includegraphics[width=0.46\textwidth]{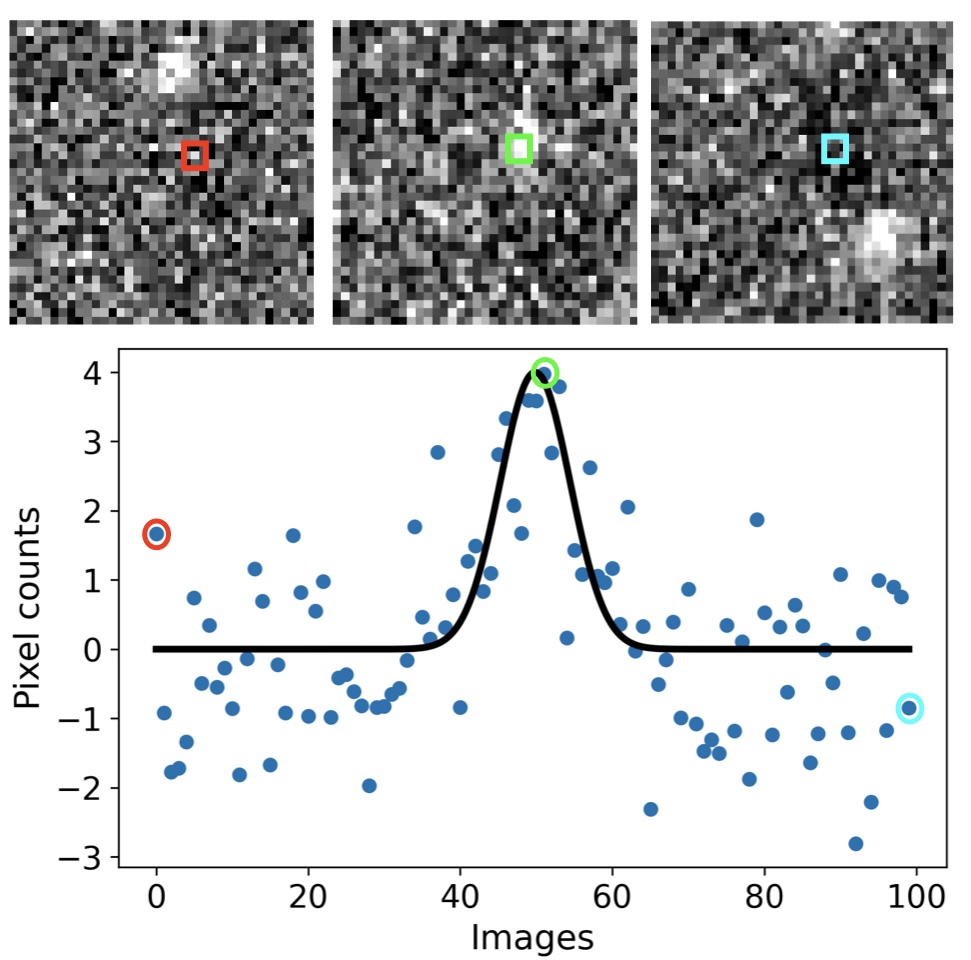}
    \caption{Upper panels show a synthetic moving object at three representative epochs (image indices 0, 50, and 99) as it traverses the image sequence. The colored boxes indicate the same detector pixel sampled at each epoch, whose values contribute to the light curve shown below. The lower panel displays the resulting single-pixel light curve, with pixel counts plotted as a function of image index. Blue points represent the full time series of pixel values, while the red, green, and cyan circled points correspond to the epochs highlighted in the upper panels. The black curve shows the expected theoretical response for a moving point source, yielding an approximately Gaussian-shaped signal as the object passes through the pixel.
    }

    \label{fig:LC}
\end{figure}

\subsection{Gaussian Motion Filter (GMoF)} \label{sub:GMoF}
The signal by a moving object over a given pixel increases as it approaches it, following a Gaussian function (black curve in Fig.~\ref{fig:LC})
with width $\sigma_{\mathrm{G}}$:
\begin{equation}
\sigma_{\mathrm{G}} = \frac{\mathrm{FWHM}/\mathrm{pixscale}}{(v / \mathrm{pixscale} / 3600) \times \Delta t}
\end{equation}
where $\mathrm{FWHM}$ is the full width at half maximum of the PSF (in arcseconds), $\mathrm{pixscale}$ is the pixel scale in arcseconds per pixel, $v$ is the apparent velocity of the object in arcsecond per hour, and $\Delta t$ is the time difference between consecutive frames in seconds. 

The GMoF takes advantage of this behavior by applying a one-dimensional Gaussian 
filter $G_{\mathrm{mov}}$ to each pixel’s time series, enhancing the signal of a moving object while suppressing background noise. As this work targets primarily Kuiper Belt Objects (KBOs), we adopted the following values:
$v = 3~\mathrm{arcsec~hr^{-1}}$ (apparent motion of KBOs), 
$\mathrm{FWHM} = 1.2~\mathrm{arcsec}$, 
$\mathrm{pixscale} = 0.2637~\mathrm{arcsec~pix^{-1}}$ (pixel scale), and 
$\Delta t = 140~\mathrm{s}$, setting $\sigma_{\mathrm{filt}} = 7$ images.
The total filter length was chosen to be approximately three times its width ($\sim3\sigma_{\mathrm{filt}}$). 

\begin{figure*}[ht!]
    \centering
    \includegraphics[width=\textwidth]{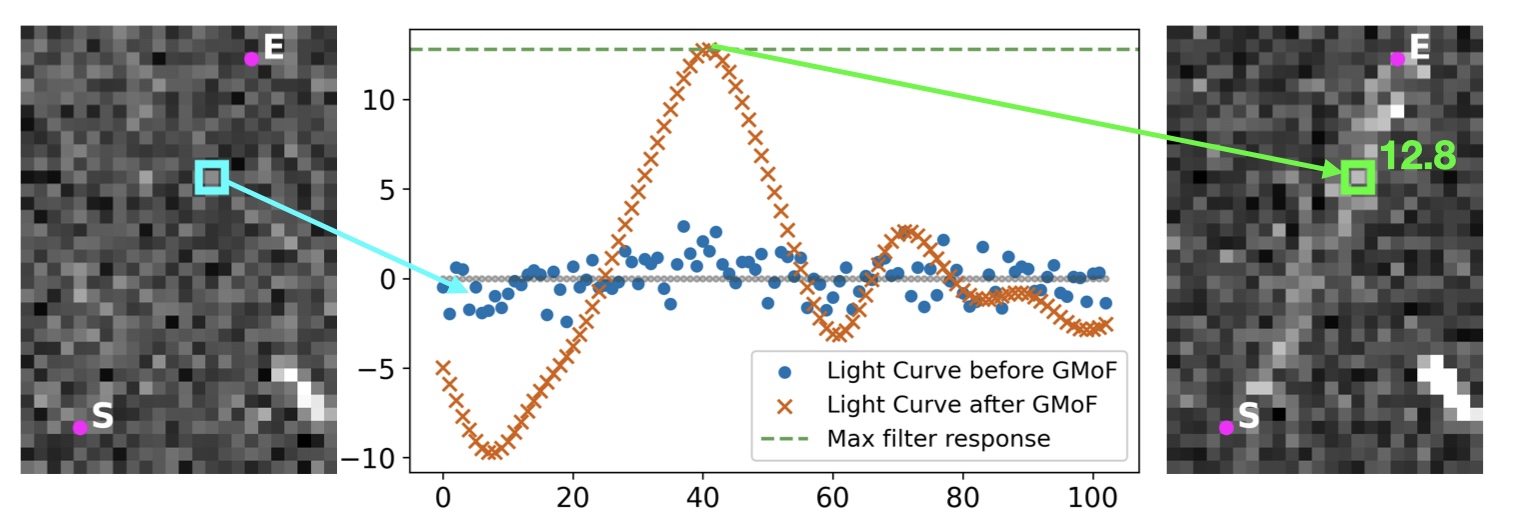}
    \caption{Comparison of two image-combination statistics applied to a planted moving object of magnitude $m=25.06$. 
    The left panel shows the coadded image, with a blue box indicating the pixel used to extract the light curve. The central panel presents the single-pixel light curve before (blue points) and after (vermilion crosses) applying the Gaussian Motion Filter (GMoF; Section~\ref{sub:GMoF}). 
    The dashed green line marks the maximum filter response. The labels S and E indicate the start and end of the object’s streak across the pixel. After applying GMoF, the enhanced signal traces the object’s passage through the pixel, producing a clear peak. The final combined image, constructed by selecting the maximum value of the filtered light curve, is shown in the right panel, where the object is recovered with enhanced significance.}

    \label{fig:GMoF-result}
\end{figure*}

Each pixel’s lightcurve, $I(t)$, is convolved with $G_{\mathrm{mov}}$, amplifying the signal associated with the object (See Fig.~\ref{fig:GMoF-result}). We adopt the maximum value of this convolved series, $\max(I(t) \ast G_{\mathrm{mov}})$, as the detection statistic. This leads to a significant improvement in the detectability of faint moving sources.
The process creates a single combined image, where we see the trails of all moving sources. 
Although in this work we only consider one $\sigma_{\mathrm{filt}}$ value, we show in the results this method is able to detect a wide range of apparent motions beyond that of distant KBOs, including Near-Earth Objects (NEOs), and Main-Belt asteroids.

\section{Machine Learning Framework and Implementation}\label{sec:ML}
The strength of deep learning lies in its ability to identify complex patterns and make rapid predictions once trained on representative datasets, as is the case for trails  in combined images.
We focused on the CNN architecture of YOLOv8-L model, the large variant of the YOLOv8 series \citep{yolov8_ultralytics}, a state-of-the-art, anchor-free object detection network released by Ultralytics on January 10, 2023. Pretrained on the Common Objects in Context (COCO) dataset \citep{lin2014microsoft}, YOLOv8-L provides robust feature representations that generalize well to new detection tasks. It offers a well-tested balance between accuracy and computational efficiency, making it suitable for detecting faint or low-contrast moving objects in astronomical images. 

\subsection{Building the training dataset} \label{sec:MLdata} 
An adequate volume of training data is essential for neural networks, as limited datasets can lead to overfitting, whereby the model performs well on the training examples but fails to generalize to unseen data. To mitigate this effect, we generated a comprehensive set of synthetic moving objects that is distinct from the sample used to construct the efficiency curve, while following the same simulation procedure described in Section~\ref{sub:preprocessing}.
The implanted objects were assigned magnitudes ranging from 19 to 27 in the VR band and apparent velocities from 3 to 5 arcseconds per hour, covering the typical range of faint and slow-moving bodies in the Solar System.

The simulated objects were implanted into three DEEP survey fields different from those listed in Table~\ref{tab:decam_fields}. These were used exclusively to construct the training, validation, and testing (TVT) datasets for the neural network, and were selected for having average seeings between (FWHM$\sim1.2$ and 1.4 arcsec). The objects in these fields present in the Minor Planet Center (MPC) database\footnote{\url{https://minorplanetcenter.net/}} 
were incorporated into the TVT dataset. The dataset also includes purely background images containing no objects, as well as images with cosmic rays, ensuring the model can discriminate between real moving sources and spurious signals. 
The resulting images were processed through the same pre-processing pipeline described in Section~\ref{sub:preprocessing}, ensuring consistency with the data used for inference.

DECam images were cropped into 512×512 tiles and contrast-normalized using the \texttt{ZScaleInterval} implementation from the Astropy visualization module, which determines intensity limits using the standard IRAF/DS9 zscale algorithm. Pixel values outside these limits were clipped, and the resulting range was linearly rescaled to 8-bit intensity values. This pre-processing emphasizes morphological features relevant for visual detection and pattern recognition by a CNN. Absolute photometric measurements are not required at this stage.
The processed tiles were stored in a compressed image format (JPEG) to reduce storage requirements and data-loading overhead.
To simulate the diversity of real object trajectories, each planted object was assigned a random orientation. Additional data augmentation, including variations in brightness, contrast, and PSF blurring, was applied to improve model robustness to different observational conditions and noise patterns. In total, the TVT dataset included over 68,420 objects distributed across 16,000 images. This complete dataset was split into training, validation, and testing sets using an 80/10/10 ratio. This results in 12,800 images for training, 1,600 images for validation, and 1,600 images for testing, which ensures that the majority of the dataset is available for model learning, while sufficient images are reserved for evaluating performance during training and for final testing to assess generalization to unseen data.

\subsection{Model training and validation}

Model training was conducted using Google Colaboratory\footnote{\url{https://colab.google/}}, a cloud-based Jupyter Notebook environment that provides ready-to-use Python libraries and free access to GPUs. The YOLOv8-L model was fine-tuned with a batch size of 16 using the Stochastic Gradient Descent (SGD) optimizer. Training was performed on an NVIDIA Tesla T4 GPU using CUDA 12.4. On-the-fly data augmentation was applied, including random translation (0.1), scaling (0.5), mosaicking (1.0), and the additional augmentations described in Section~\ref{sec:MLdata}. The model was trained for approximately 200 epochs over $\sim$46 hours of GPU time, after which the training and validation losses showed stable convergence. This resulted in a robust object-detection model capable of reliably identifying faint moving-object trails in astronomical images.

To quantify performance, we used the mean Average Precision (mAP) metric, which summarizes the trade-off between precision and recall. The Average Precision (AP) for a single class corresponds to the area under the precision–recall curve. 
Model predictions were categorized into three outcome types using the Intersection-over-Union (IoU) criterion. 
A detection was classified as a True Positive (TP) when the predicted bounding box correctly matched a ground-truth object with $\mathrm{IoU} \ge 0.5$. A False Positive (FP) was assigned when a predicted box did not correspond to any real object (i.e., $\mathrm{IoU} < 0.5$) or when it duplicated an existing detection. A False Negative (FN) occurred when a real moving object in the image was not detected at all (no prediction with $\mathrm{IoU} \ge 0.5$). We also consider the AP when the IoU is 0.5 and 0.95 as AP$_{50}$ and AP$_{95}$ respectively. We compute precision as $TP/(TP+FP)$ and recall as $TP/(TP+FN)$.

Training evaluation followed the COCO-style metrics adopted by Ultralytics\footnote{\url{https://docs.ultralytics.com/guides/yolo-performance-metrics/}}. 
The final model achieved a precision of 0.80, a recall of 0.71, AP$_{50}$=0.77, and AP$_{95}$=0.57 on the validation set. These results demonstrate strong performance and a good balance between accurate localization and completeness in detecting faint moving objects across varying background noise levels. When visually inspecting these False Positive detections are dominated by residual image artifacts, imperfect background subtraction near bright sources, or chance noise alignments that mimic elongated trails in the motion-filtered stack.

\subsection{Model deployment}
\begin{figure}[ht!]
\centering
\gridline{
  \fig{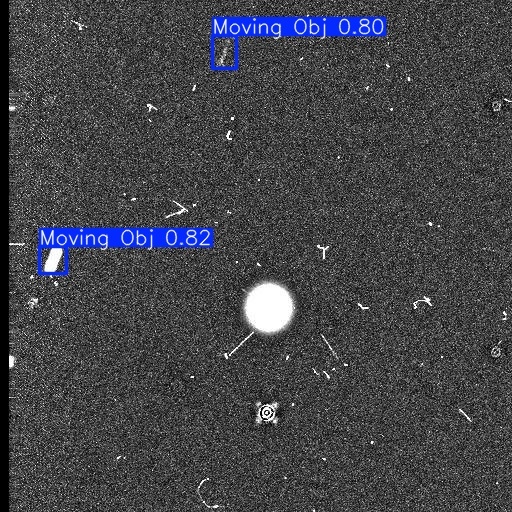}{0.23\textwidth}{(a)}
  \fig{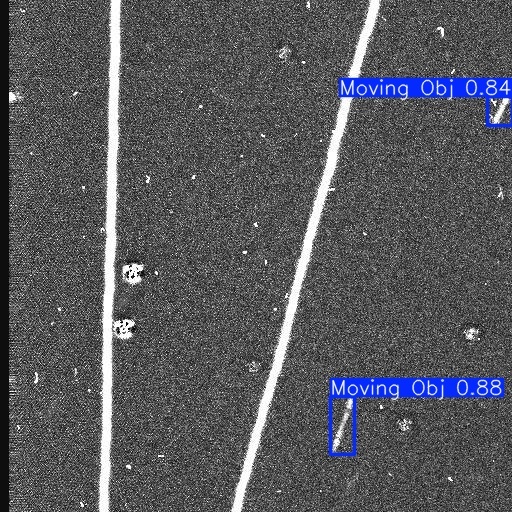}{0.23\textwidth}{(b)}
}
\caption{Detections by our best-trained model
marked with blue boxes labeled “Moving Obj,” and the 
model’s detection confidence. Poorly subtracted stars, image artifacts, and cosmic rays were successfully rejected. Panel (b) also contains two satellite trails, which the model correctly ignored.
}
\label{fig:yolo_output}
\end{figure}

When the model was fine-tuned on the custom dataset of slow-moving Solar System objects, best-performing trained model, which was subsequently deployed on the stacked images of the field-night dataset listed in Table~\ref{tab:decam_fields}, comprising a total of 7808 images. The model requires less than 11~milliseconds to generate predictions for each image and outputs both annotated images with detection confidence scores and accompanying text files containing the localization vectors of the predicted objects. These text files follow the same format as the input annotations. As illustrated in Fig.~\ref{fig:yolo_output}, the model demonstrates strong performance in detecting slow-moving objects while effectively ignoring artifacts, poorly subtracted stars, cosmic rays and also satellite trails. 

\subsection{Post YOLO predictions} \label{sub:post-yolo}
\begin{figure}[htbp]
\centering
\gridline{
  \fig{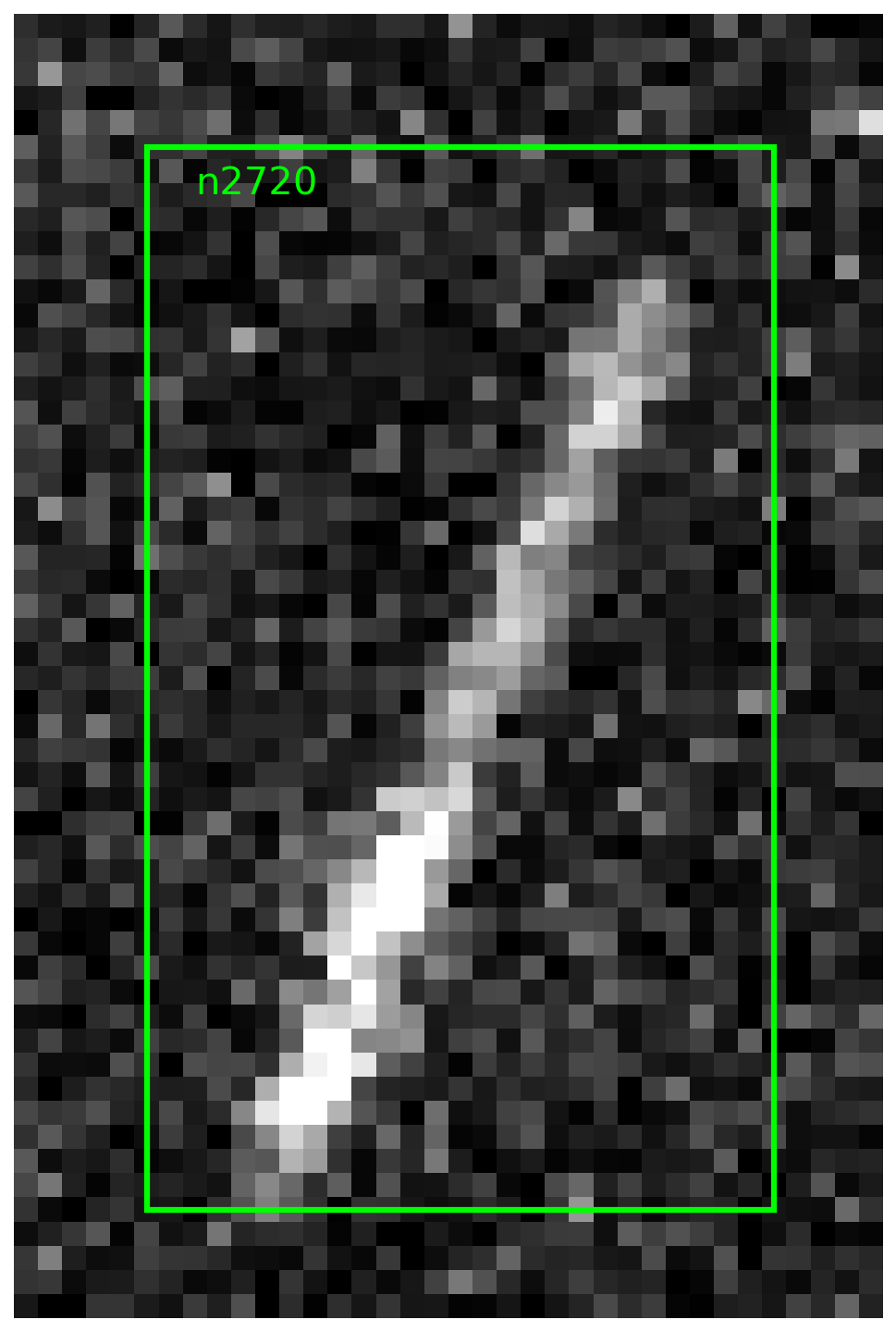}{0.226\textwidth}{(a)}
  \fig{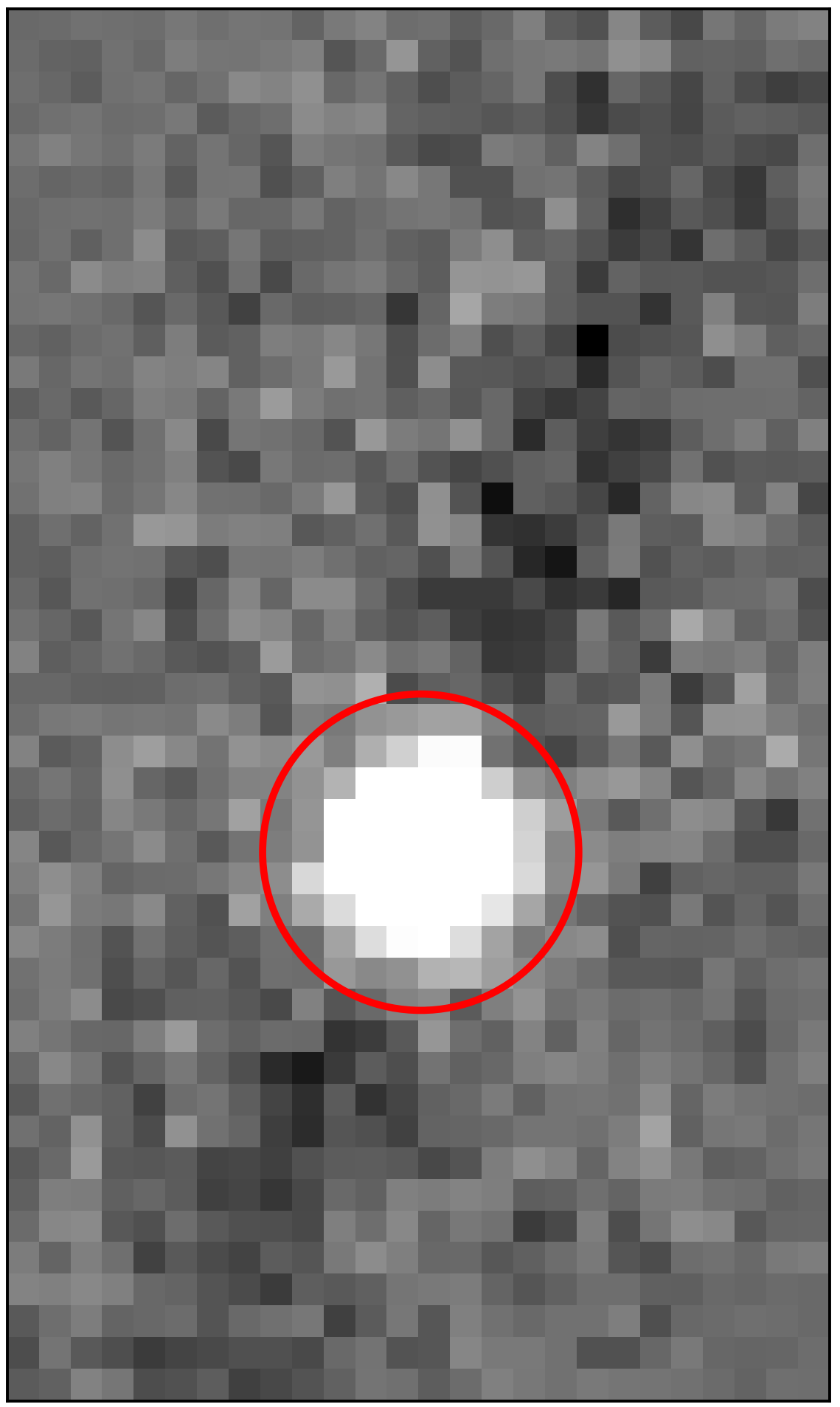}{0.20\textwidth}{(b)}
}
\vspace{-1mm}
\gridline{
  \fig{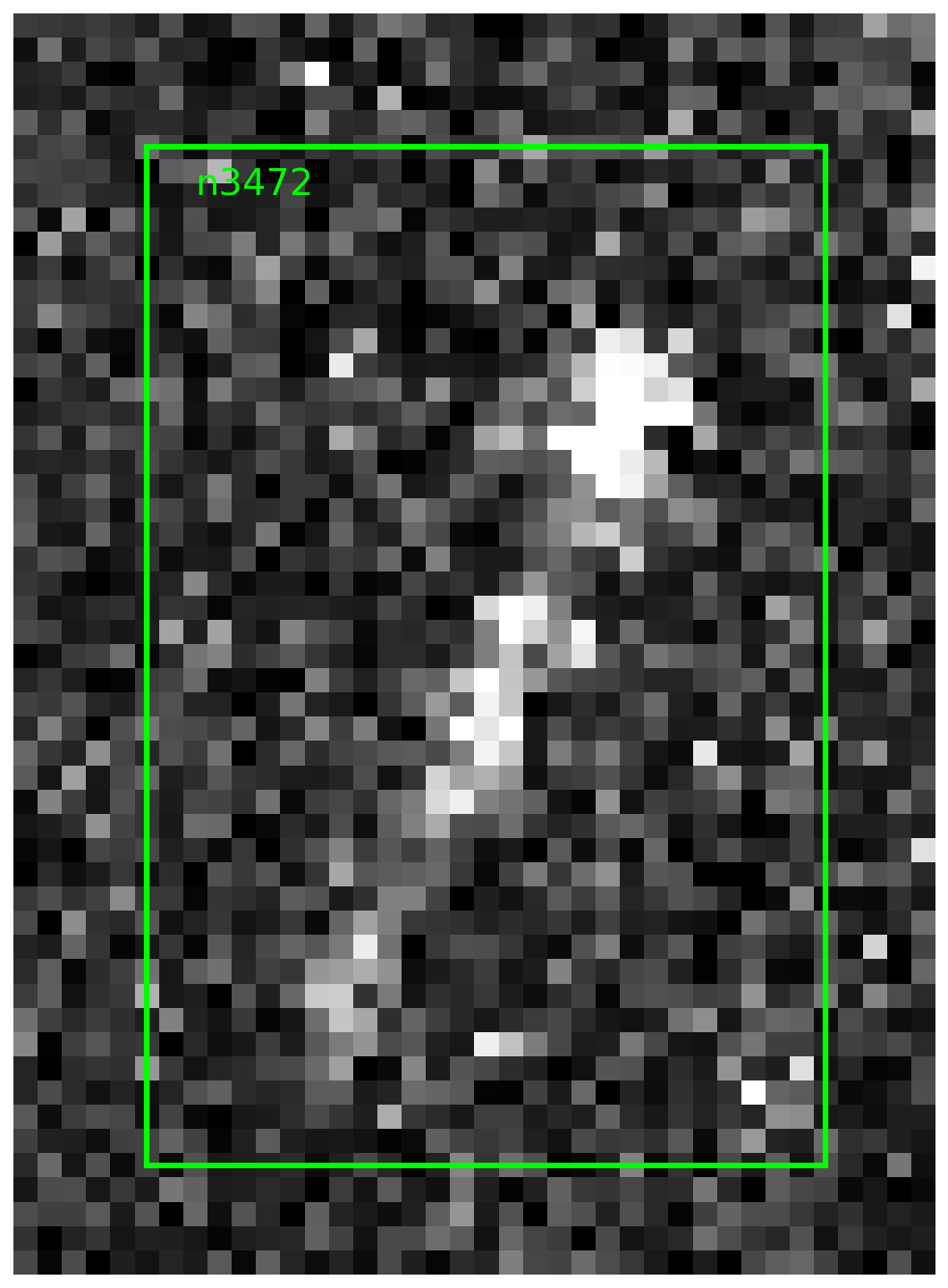}{0.23\textwidth}{(c)}
  \fig{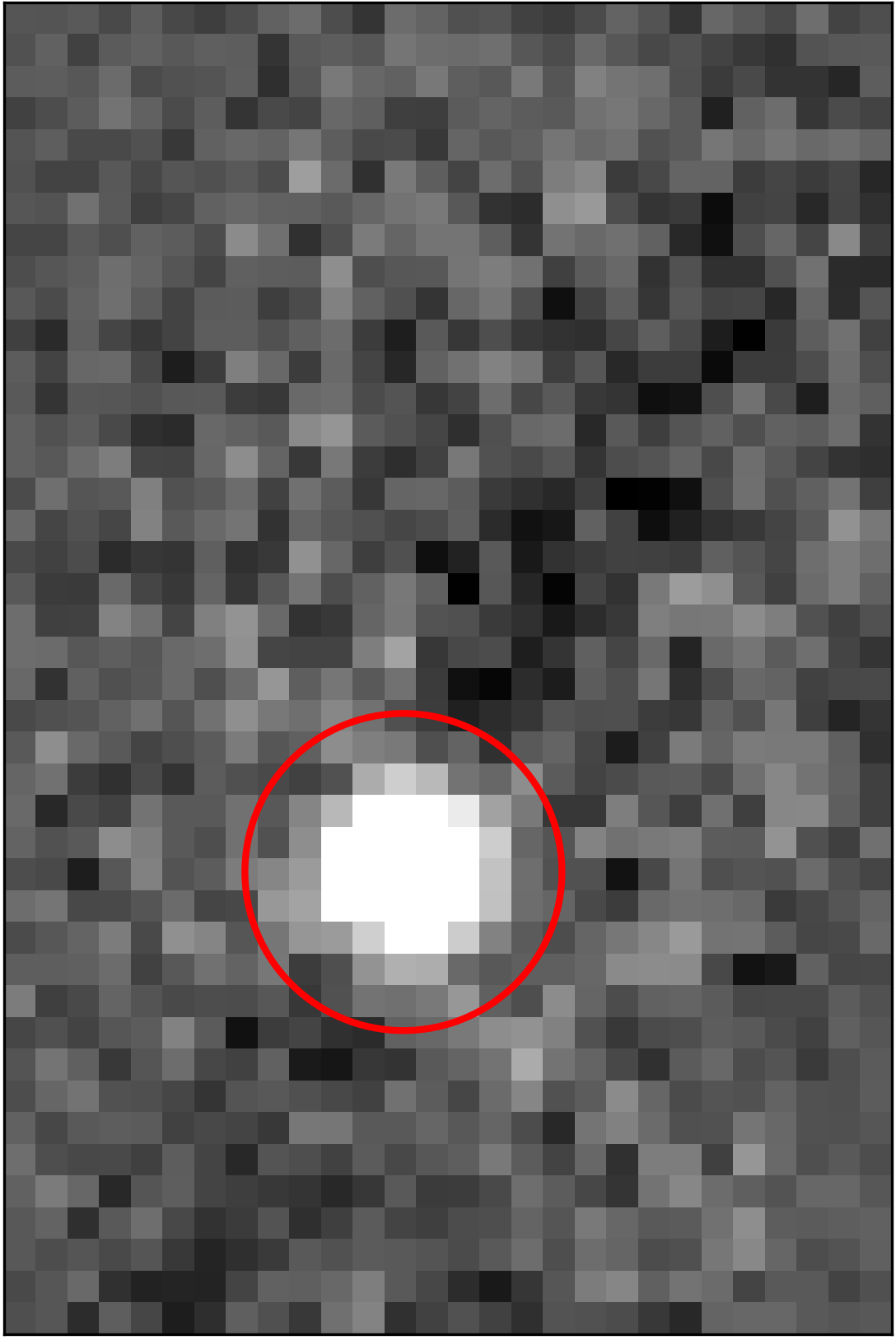}{0.21\textwidth}{(d)}
}

\caption{
Panels (a) and (c) show the objects flagged by the machine learning model \texttt{n2720} and \texttt{n3472}. Panels (b) and (d) show these confirmed detections after shifting and stacking to their right rate of motion, now renamed as \texttt{B1j\_YOSO\_04} (m$\sim$24.3) and \texttt{B1j\_YOSO\_08} (m$\sim$24.9), respectively.} 
\label{fig:new}
\end{figure}

From the CNN we receive bounding boxes of the objects, which contain the end points of their trails, as seen in the left images in the Fig. \ref{fig:new} the trails of the objects \texttt{n2720} and \texttt{n3472} surrounding by the green bounding box predicted by our best trained model. While the bounding box localizes the trail, it does not uniquely determine the direction or rate of motion. The trail lies along a diagonal of the box, and the object may be moving in either of the two opposite directions along that diagonal.

To resolve this ambiguity and confirm whether a detected object is a genuine Solar System moving object, the pipeline performs a bi-directional shift-and-stack procedure 
during the refinement stage. We shift each candidate box along its two diagonals, evaluating both directions along each diagonal. Each resulting stacked image is analyzed using the Source Extraction and Photometry (SEP) library \citep{bertin1996sextractor}, a Python implementation of the widely used \texttt{SExtractor} algorithm, to extract sources and evaluate their shapes. SEP characterizes the shape of a detected source by measuring the second moments of its flux distribution relative to the source centroid. From these moments, the algorithm derives the parameters of an equivalent ellipse describing the light distribution, including the semi-major ($a$) and semi-minor ($b$) axes. The ellipticity of the source is then defined as $e = 1 - b/a$, where values close to zero correspond to a round, point-like object and larger values indicate an elongated source.

The pipeline iteratively refines the shift rates used in the stacking procedure. After each stack, SEP is applied to the combined image to measure the source shape parameters. Candidate shift rates are evaluated by minimizing the measured ellipticity while also favoring solutions that produce higher flux and a source centrality in the image. This procedure searches for the shift rate that produces the most compact and round stacked source, corresponding to the correct apparent motion of the object. Optionally, a Nelder–Mead optimization is applied to further refine the velocities, penalizing low-flux or off-center objects. The right panels of Fig.~\ref{fig:new} show the final stacked images for \texttt{n2720} and \texttt{n3472}, renamed as \texttt{B1j\_YOSO\_04} and \texttt{B1j\_YOSO\_08} by the author, illustrating the outcome of this optimization process.

This stage serves as an automated vetting step. 
Candidates are retained only if, in the final shift-and-stack image, they are detected 
above a $1.5\sigma$ threshold relative to the local background, and exhibit circular morphology with FWHM~$\geq 2.5$ and a very low 
ellipticity ($e = 1 - b/a < 0.001$).

\section{Photometry}
\subsection{Magnitude Determination from Stacked Images}
\label{sec:photometry}
For each detected moving object, photometry is performed on the stacked image produced by the YOSO pipeline. The stacking process combines flux from multiple individual exposures, each of which may have a different photometric zero-point due to variations in observing conditions and instrumental response. As a result, the object magnitude cannot be derived using a single effective zero-point without loss of accuracy. 

\begin{figure}[ht!]
\plotone{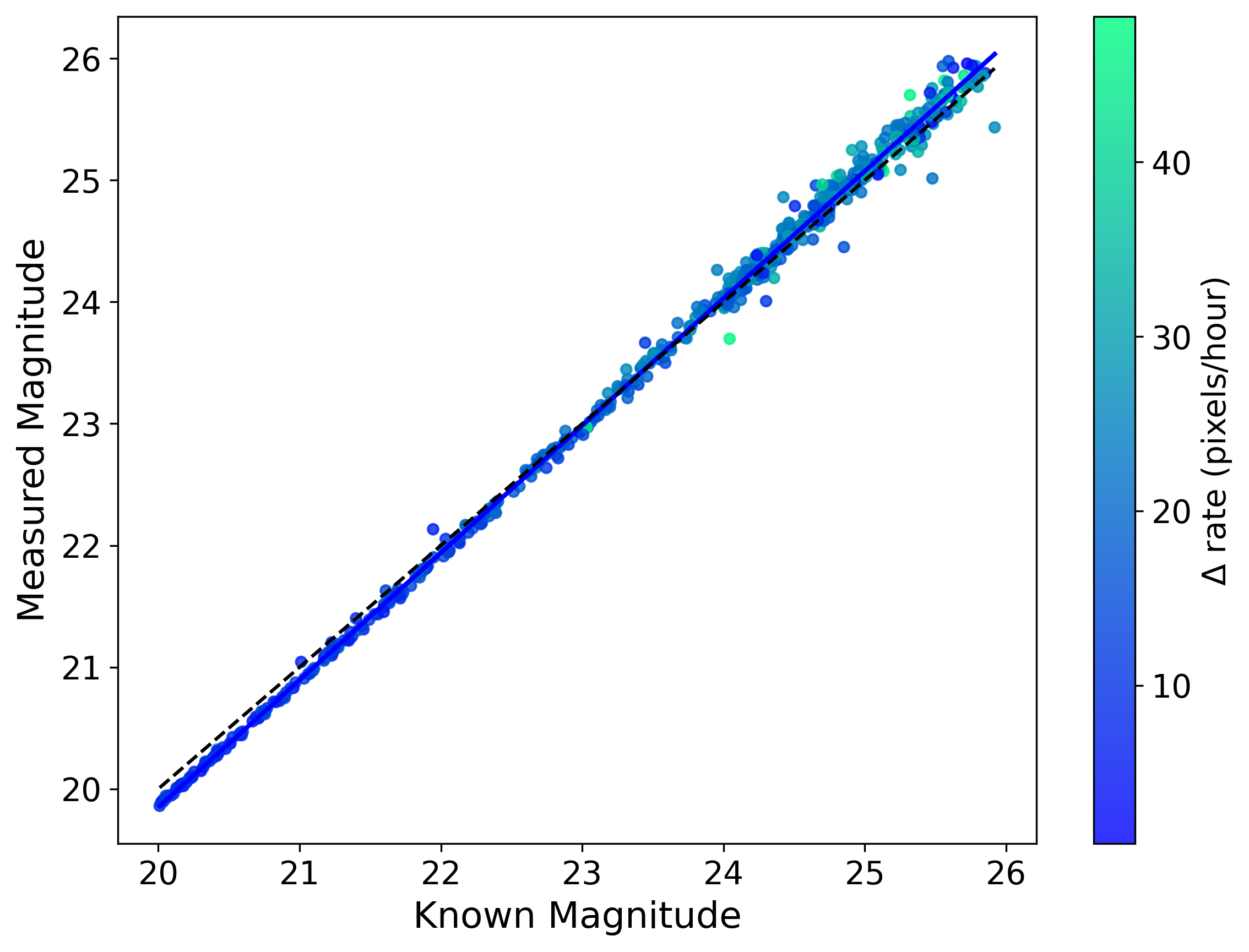}
\caption{Comparison between implanted magnitudes for synthetic objects their measured photometry. 
Points are color-coded by their motion rate (pixels/hour). The blue solid line shows the best-fit linear calibration, and the dashed black line represents a one-to-one relationship.
\label{fig:phot_fit}}
\end{figure}

For every object we first compute the photometric zero-point of the stacked image by considering that of each of the CCDs that went into it. When comparing our photometry with the implanted magnitudes we get a mean error of $0.14$ magnitudes and a standard deviation of $0.33$ magnitudes 
(Fig.~\ref{fig:phot_fit}).

\subsection{Magnitude Calibration} \label{mag_cal}

We calibrated the instrumental magnitudes onto the reference system by fitting a linear relation between measured and catalog magnitudes for a set of calibration objects. Residuals are defined as the difference between the calibrated magnitudes and the catalog values.

For our sample of $N = 567$ synthetic moving objects, our photometry yields residuals with
an RMS scatter of $0.08$~mag. This narrow, nearly unbiased distribution indicates that the calibration is accurate at the $\sim 0.1$~mag level 
as plotted in Fig.~\ref{fig:phot_fit}.

\section{Results} \label{results}
We tested the detection efficiency of our method by implanting synthetic objects into the same images, as described in Section \ref{sec:method}. After analyzing four field nights from the 
$B1$ quadrant, we obtained the following results.

\subsection{Limiting Magnitudes and Recovery}

\begin{figure}[ht!]
\plotone{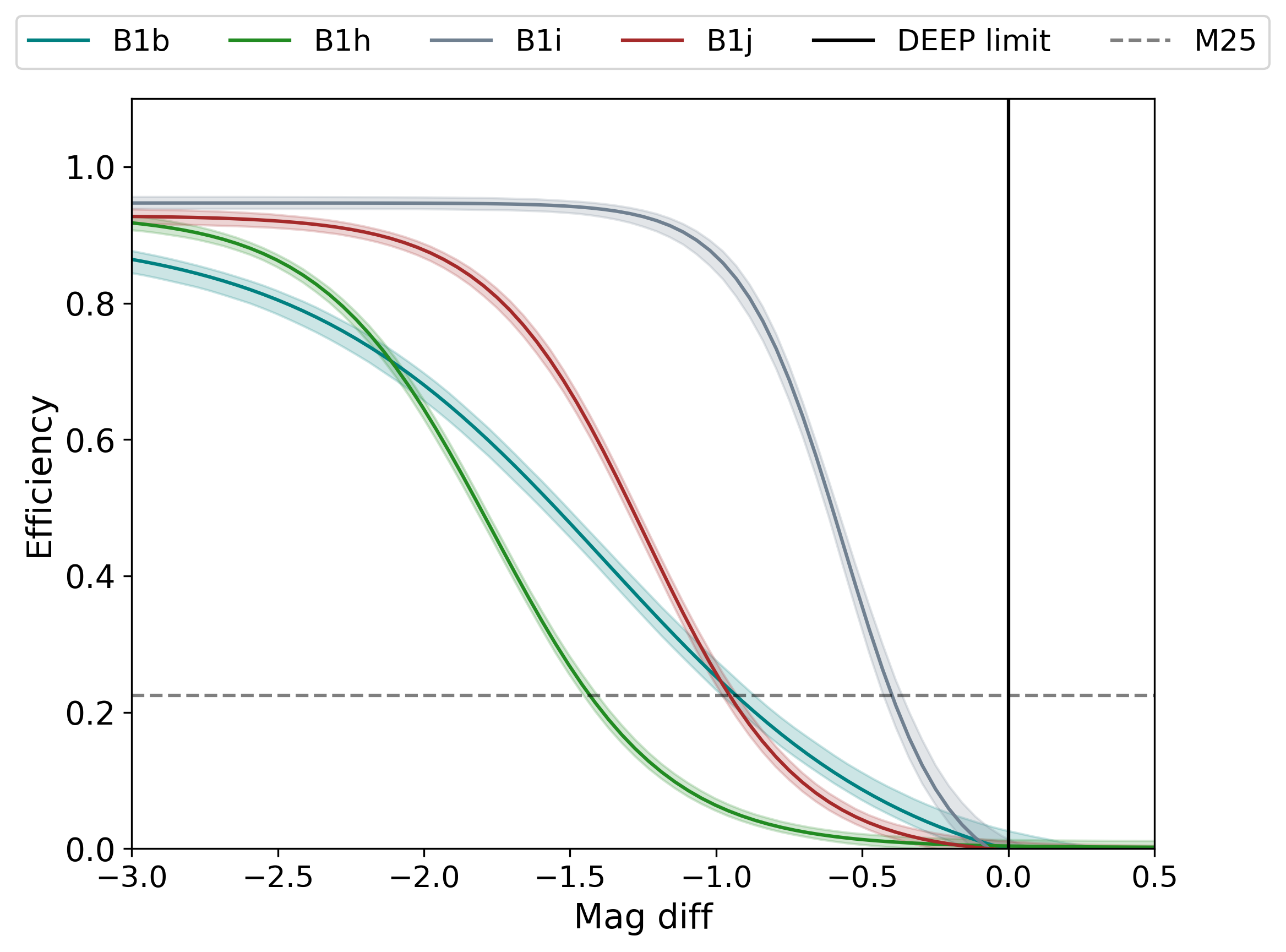}
\caption{Detection efficiency as a function of magnitude difference relative to the DEEP survey limit for each analyzed field night. Colored curves show the recovery efficiency obtained with our method, while the shaded regions indicate the associated uncertainties. The black vertical line at $\Delta m = 0$ marks the nominal DEEP detection limit reported by \cite{napier2024decam}. The horizontal dashed line denotes the 25$\%$ recovery efficiency threshold. Offsets to the left of the vertical line correspond to brighter sources, and the displacement of each curve relative to this reference illustrates how much shallower each field night is compared to the DEEP survey. 
\label{fig:eff_plot}}
\end{figure}

The results of reanalyzing the four field nights are summarized in Fig.~\ref{fig:eff_plot}.  The detection efficiency curves and limiting magnitudes shown in this figure are derived 
from injection-and-recovery tests using synthetic objects, as described in Section~\ref{sec:method} and Section \ref{sec:ML}. We note that the post-YOLO verification procedure described in Section~\ref{sub:post-yolo} reveals a similar precision for the YOLO results (0.7) to the one obtained in the training and testing of the model. 

These synthetic sources provide a controlled and uniform sample for quantifying the recovery efficiency of our pipeline as a function of magnitude. Based on these tests, our method reaches a limiting magnitude that is on average $\sim$0.88~mag brighter than that achieved by KBMOD. 
Detailed results for each field night are provided in Table~\ref{tab:eff_value}.

\begin{figure}[ht!]

        \centering
        \plotone{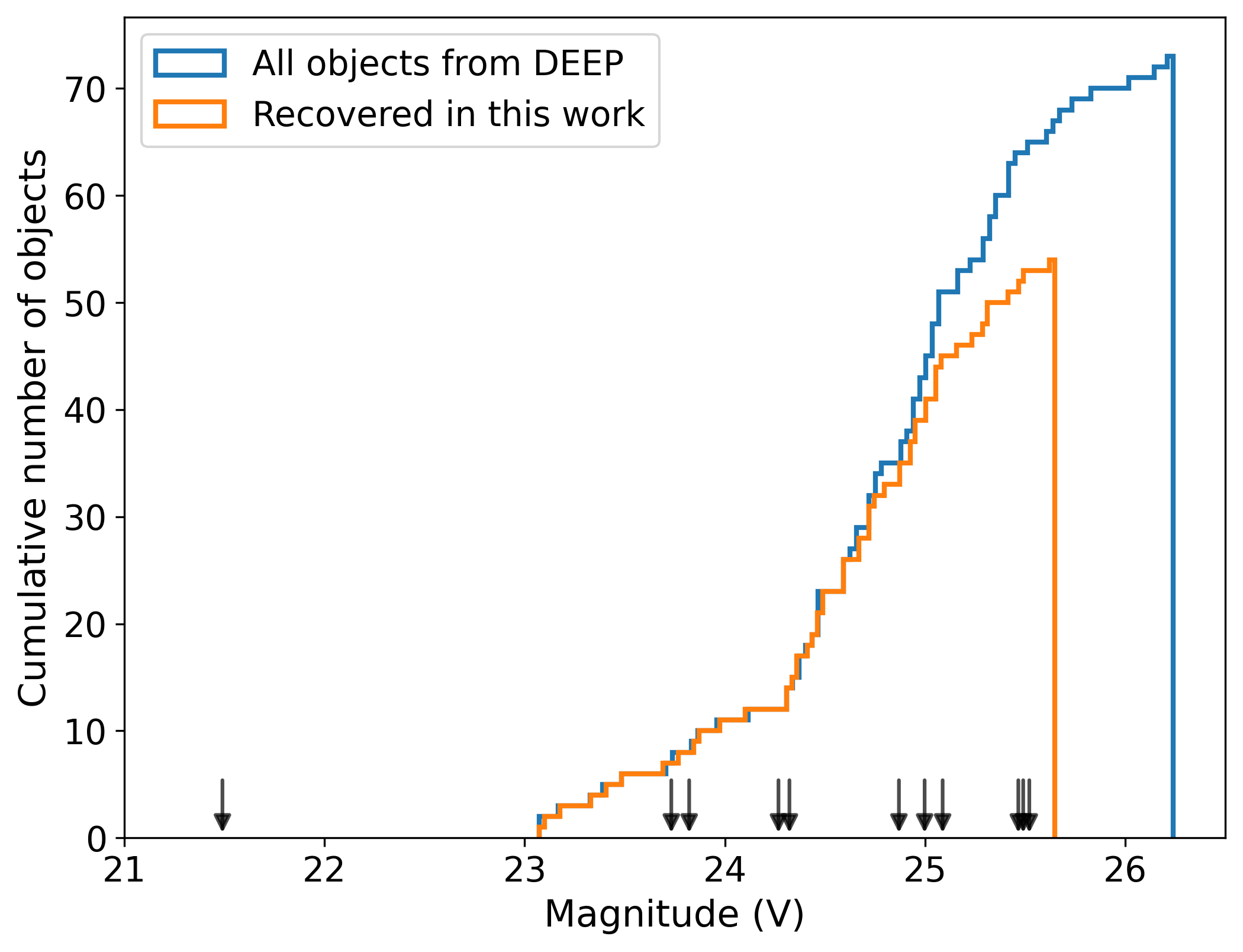}
        \caption{
        Cumulative magnitude distribution of Trans-Neptunian Objects in the analyzed DEEP survey fields \ref{fig:FN}. 
        The blue step curve shows all TNOs reported by the DEEP survey \citep{smotherman2024decam}, while the orange step curve shows the subset of those objects recovered by our pipeline. 
        Downward-pointing arrows mark the calibrated magnitudes of newly discovered TNOs identified by our method that were not reported in the DEEP catalog, magnitudes are computed using the photometric calibration described in Section~\ref{mag_cal}.
        }
    \label{fig:RF}
\end{figure}

Using the empirically determined detection limit from the synthetic injection-and-recovery tests, we then compare our results to the catalog of real TNOs reported by \citet{smotherman2024decam}. 
As illustrated by the cumulative magnitude distributions in Fig.~\ref{fig:RF}, we do not recover any of the previously reported TNOs fainter than $m \simeq 25.6$. 
This behavior is fully consistent with the limiting magnitude inferred from the synthetic-object recovery analysis shown in Fig.~\ref{fig:eff_plot}.

\begin{table}[ht]
    \centering
    \begin{tabular}{|c c c c c|} 
     \hline
     Field & Night & KBMOD & This work & Recovery (\%) \\ 
     \hline
     B1b & 15-10-2020 & 26.05 & 25.15 & 66 \\ 
     
     B1h & 08-09-2021 & 25.55 & 24.65 & 45 \\
     
     B1i & 03-10-2021 & 26.34 & 26.04 & 70 \\
     
     B1j & 30-09-2021 & 25.74 & 24.84 & 50 \\ 
     \hline
    \end{tabular}
    \caption{Comparison of the achieved limiting magnitude between \cite{napier2024decam} (using KBMOD; see \cite{whidden2019fast}) and this work, together with the fraction of object recovered by YOSO pipeline. Over the four field nights, KBMOD detected 73 TNOs, of which 45 were re-identified by our method. The recovery column provides the ratio of objects recovered here relative to the original KBMOD detections.}
    \label{tab:eff_value}
\end{table}

\subsection{New Detections}
Even though our current analysis does not reach the same depth as KBMOD, our pipeline has identified additional TNO candidates that were not included in \cite{smotherman2024decam}. Across the four analyzed field nights, we have detected 11 new TNOs. For comparison, these objects are marked on the cumulative magnitude distribution in Fig.~\ref{fig:RF}, each new detection is indicated by a downward-pointing black arrow at the corresponding magnitude. 

Notably, one of these objects is relatively bright (m = 21.49), brighter than lowest detection limit in the cumulative magnitude distribution. However, this can be explained by the intrinsic recovery statistics of KBMOD, while KBMOD detects essentially all objects brighter than its nominal limiting magnitude, it recovers only a fraction
of the full population due to selection effects and internal detection thresholds. 

\begin{figure*}[ht!]
    \centering
    \includegraphics[width=\textwidth]{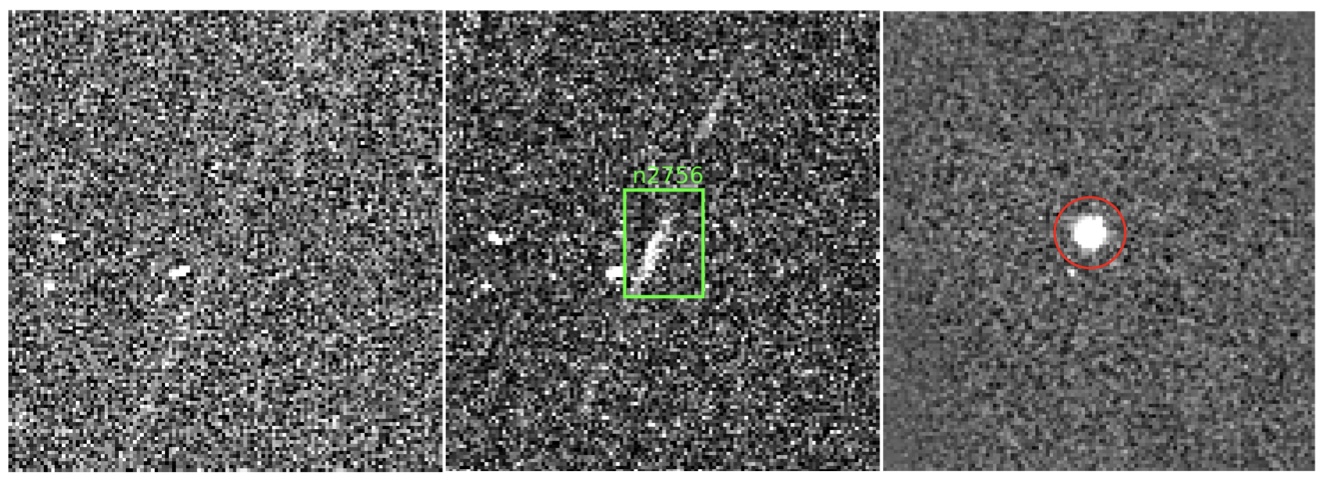}
    \caption{(Left) A newly detected object ($m \sim 20.5$), not previously reported in the DEEP catalog, is barely noticeable in the average image. Based on its apparent motion, it is consistent with a main-belt asteroid.
  (Middle) The motion-filtered image enhances its trail, and YOSO flags a segment  of it as \texttt{n2756}. (Right) The shift-and-stack refinement converges to a precise apparent motion, of $(\dot{\alpha}, \dot{\delta}) \approx(-26.00, -10.42)$ arcsec hr$^{-1}$, corresponding to a total sky-plane motion of $\sim$28 arcsec hr$^{-1}$.}

    \label{fig:ast}
\end{figure*}

In addition to the TNO search, YOSO identified 216 previously unreported faster-moving objects with apparent sky-plane rates between 10 and 60~arcsec~hr$^{-1}$, as well as 27 with rates exceeding 60~arcsec~hr$^{-1}$. We note that the study of \citet{smotherman2024decam} was not optimized for the detection of fast-moving objects, and this comparison should therefore be interpreted in that context. All detections reported here correspond to single-night identifications. Notably, all new discoveries, including new TNOs, were obtained using a single velocity hypothesis in the GMoF (\ref{sub:GMoF}), corresponding to an assumed rate of 3~arcsec~hr$^{-1}$, $\sigma_{filt}=7$. Despite this single-velocity assumption, the method successfully recovered objects spanning a wide range of apparent motions, from distant TNOs to main-belt asteroids and near-Earth objects.  In Fig.~\ref{fig:ast} we show how our machine learning method flagged an object that happened to be a segment of a probable asteroids' trail. The right panel shows the result of YOSO after finding the right rate of motion using the shift-and-stack procedure in the refinement stage. Although the flagged detection in this figure is only a portion of the asteroid's track, the shift and stack procedure converges only when the signal from the moving source reaches a minimum in ellipticity along the apparent velocity vector.

\section{Discussion} \label{discussion}
Across the four field nights analyzed here, YOSO recovers 45 of the 73 TNOs reported by \citet{smotherman2024decam} and detects 11 additional TNOs. KBMOD is explicitly tuned to maximize sensitivity at the faintest flux levels through a dense grid of trial rates and directions at the cost of generating a large number of candidate trajectories that require aggressive vetting. In contrast, YOSO compresses the full temporal information into a single motion-filtered stack and relies on a machine-learning classifier. 
YOSO has a final step in which a shift-and-stack algorithm improves the initial rate of motion based on its discovery box. This step leaves only high-confidence trails that when shifted are consistent with a point source. This leaves a purity rate for YOSO close to 99\%.

The TNOs detected by YOSO but not reported by \citet{smotherman2024decam} reflect the differences between the two approaches. In particular, YOSO is able to recover objects that are partially masked or located in regions of complex background, and that would appear noisy in a KBMOD stack. These results indicate that combining independent pipelines with distinct systematics can increase the overall completeness of TNO searches in deep imaging surveys.

A central design choice in this implementation of YOSO is the use of a single characteristic rate of motion in the GMoF. While this assumption does not match the full range of apparent motions in the data, it still manages to enhance objects over a broad velocity range. One direct improvement to the depth of YOSO would be to try different rates for the GMoF, at the cost of extra combined images to search. This would still require fewer trials than a traditional shift-and-stack method since only the rate is involved, not the direction. 

An additional strength of YOSO is its sensitivity to fast-moving objects. Although the primary focus of this work is the detection of distant, slow-moving TNOs, the same pipeline identifies more than 200 previously unreported NEOs and main-belt asteroids in the analyzed fields. These detections arise because fast-moving sources also produce high-contrast, elongated features in the motion-filtered stack, which are readily identified by the trail-based classifier. This demonstrates that a single, unified framework can simultaneously probe a wide range of apparent motions. It is noteworthy that $\sigma_{filt}\rightarrow 1$ for objects that move faster, meaning that at some point YOSO could be used in data with sparse cadence, like the LSST Wide Fast Deep (WFD) \footnote{\url{https://survey-strategy.lsst.io/baseline/wfd.html}}.

Our goal in this work is to develop and characterize the detection stage of a moving-object pipeline. We therefore restrict the analysis to single-night detections and do not attempt multi-night or multi-year linking or orbit determination. These tasks are typically handled by dedicated survey pipelines and can, in principle, be applied to YOSO candidate lists as a follow-up stage without modification to the detection framework.

In addition to differences in depth and selection function, YOSO offers significant computational advantages compared to traditional shift-and-stack pipelines such as KBMOD. The first, a single combined image versus a dense grid of trial rates and directions. Also, although training a neural network represents a one-time computational expense, inference on new data is extremely fast. 
This computational profile makes YOSO well suited for large-scale surveys and repeated re-analysis like LSST, and space-based missions where processing is constrained and data transmission is limited. 
 
In the next section, we discuss how these characteristics make YOSO a promising tool for upcoming wide-field surveys and a variety of time-domain applications.

\section{Conclusion}
We present You Only Stack Once (YOSO), a new framework for detecting faint moving objects in time-series astronomical imaging data. Our method is both simple and robust, making it easily deployable on large astronomical datasets. By utilizing a single combination of all registered images, it efficiently detects moving objects through an off-the-shelf AI-based algorithm. The approach substantially reduces false positives compared to traditional methods, providing a reliable and efficient alternative for identifying objects fainter than those detectable in individual images. This makes it a strong complement to established techniques such as shift-and-stack.

This framework opens multiple opportunities for future applications. When extended to faster-moving populations such as asteroids and Near-Earth Objects (NEOs), it could significantly improve the detection efficiency of small, faint NEOs in large surveys like LSST, particularly in deep-drilling fields. The relevance of YOSO to LSST Solar System science is strongly shaped by both observing cadence and survey footprint\footnote{\url{https://survey-strategy.lsst.io/baseline/index.html}}. LSST’s Deep Drilling Fields provide dense intra-night sampling, often consisting of tens to hundreds of short exposures obtained over several hours, a regime that is particularly well matched to YOSO’s single-stack, motion-enhancement strategy. Among the planned deep drilling fields, COSMOS lies close to the ecliptic plane and is therefore expected to yield the highest surface density of Solar System objects, making it especially favorable for faint Trans-Neptunian Objects and small near-Earth objects. XMM–LSS, located at moderate ecliptic latitude, offers sensitivity to dynamically excited and higher-inclination populations. In contrast, the Wide-Fast-Deep survey, with its sparser intra-night cadence, is less optimal for ultra-faint slow movers but remains well suited for the detection of fast-moving asteroids and near-Earth objects in single-night data. In this context, YOSO naturally complements LSST’s cadence-driven discovery space by enabling high-purity detections in densely sampled observations without the computational overhead of traditional velocity-space searches.

Space-based observatories such as the upcoming NEO Surveyor mission \cite{mainzer2023near} could greatly benefit from incorporating onboard image combination. By transmitting a single stacked frame per Visit instead of multiple raw exposures, the overall data volume and bandwidth requirements could be substantially reduced. Moreover, performing onboard stacking would enable real-time or near-real-time detection of bright, fast-moving near-Earth objects (NEOs), where a lightweight AI model could autonomously flag high-priority detections for immediate downlink and follow-up.

Beyond the detection of slow-moving Solar System objects, our approach has broader applicability in areas where maximizing the signal-to-noise ratio is essential. In time-domain astronomy, for example, the GMoF technique could be adapted to identify subtle brightness variations in variable stars or the early brightening phases of supernova progenitors by stacking lightcurves. Enhancing such sub-threshold signals may allow the detection of low-amplitude variability or pre-explosion activity that remains undetectable in individual exposures. Likewise, optimized stacking and filtering strategies could aid in the search for optical counterparts of gravitational-wave events, where combining multiple short exposures increases the likelihood of recovering faint, transient sources buried in the noise.  This method also enables the combination of ultra-deep fields that require subtraction of natural and artificial moving objects. By flagging pixels affected by objects too faint to be detected in a single image one can construct a superior source mask.

With suitable adaptation, this approach can also be extended to Angular Differential Imaging (ADI) datasets for exoplanet detection in young planetary systems, which face similar challenges of limited detection sensitivity and high false-positive rates due to quasi-static speckles from atmospheric or instrumental aberrations not fully corrected by adaptive optics \citep{follette2023introduction}. In ADI, exoplanets trace rotational trajectories defined by the changing parallactic angle, while the stellar halo remains quasi-static. The built-in subtraction module in YOSO (ISIS) can be adapted to minimize self-subtraction and preserve faint off-axis planetary signals, enabling more reliable discrimination between genuine companions and residual speckles near the host star. Preliminary tests on VLT/SPHERE observations of the HIP 70890 system show promising detections of companions at 2.1, 3.8, and 4.47 AU, clearly distinguished from the surrounding speckle noise.

In general any experiment where small particles or tracers move across sequential images would benefit from trying YOSO in order to detect low SNR signals. Examples beyond astronomy include fluid dynamics, plasma physics, and satellite tracking.

\section*{Acknowledgments}
This project makes use of data obtained with the Dark Energy Camera (DECam), designed and built by the Dark Energy Survey (DES) collaboration. The DES has been supported by the U.S. Department of Energy, the U.S. National Science Foundation, the Ministry of Science and Education of Spain, the Science and Technology Facilities Council of the United Kingdom, the Higher Education Funding Council for England, the National Center for Supercomputing Applications at the University of Illinois at Urbana–Champaign, the Kavli Institute for Cosmological Physics at the University of Chicago, the Center for Cosmology and Astro-Particle Physics at the Ohio State University, the Mitchell Institute for Fundamental Physics and Astronomy at Texas A\&M University, Financiadora de Estudos e Projetos, Fundação Carlos Chagas Filho de Amparo à Pesquisa do Estado do Rio de Janeiro, Conselho Nacional de Desenvolvimento Científico e Tecnológico, the Ministério da Ciência, Tecnologia e Inovação, the Deutsche Forschungsgemeinschaft, and the various collaborating institutions within DES.

The DES collaborating institutions include: Argonne National Laboratory; University of California, Santa Cruz; University of Cambridge; Centro de Investigaciones Enérgeticas, Medioambientales y Tecnológicas (Madrid); University of Chicago; University College London; the DES-Brazil Consortium; University of Edinburgh; ETH Zürich; Fermi National Accelerator Laboratory; University of Illinois at Urbana–Champaign; Institut de Ciències de l’Espai (IEEC/CSIC); Institut de Física d’Altes Energies; Lawrence Berkeley National Laboratory; Ludwig-Maximilians-Universität München and the associated Excellence Cluster Universe; University of Michigan; NSF NOIRLab; University of Nottingham; Ohio State University; the OzDES Consortium; University of Pennsylvania; University of Portsmouth; SLAC National Accelerator Laboratory; Stanford University; University of Sussex; and Texas A\&M University.
The observations used in this work were obtained at the Cerro Tololo Inter-American Observatory (CTIO), operated by NSF’s NOIRLab under AURA’s management through a cooperative agreement with the National Science Foundation (Proposal ID 2019A-0337; PI: D. Trilling).
This research was supported by the BASAL Center for Astrophysics and Associated Technologies (CATA), project ANID BASAL FB210003, and ANID National scholarship file number 21242203.

\facilities{Blanco (DECam), NOIRLab Astro Data Archive, DECam Community Pipeline \cite{decam-pipeline}, JPL Horizons, Minor Planet Center (MPC) database.}

\software{astropy \citep{astropy:2013,astropy:2018,astropy:2022},
    numpy \citep{numpy},
    scipy \citep{scipy},
    pandas \citep{pandas},
    matplotlib \citep{matplotlib},
    photutils \citep{larry_bradley_2021_5796924},
    ultralytics/YOLOv8 \citep{yolov8_ultralytics},
    Google Colaboratory  \citep{googlecolab},
    SAOImage DS9,
    NASA ADS. 
}

\bibliographystyle{aasjournal}
\bibliography{bibliography}{}

\end{document}